\documentclass[prx,aps,preprint,superscriptaddress,preprintnumbers,endfloats, longbibliography
]{revtex4-1}
\usepackage{graphicx}
\usepackage{dcolumn}
\usepackage{bm}
\usepackage{color}
\usepackage{sidecap}
\usepackage{amssymb}

\begin{document}

\title {A directly phase-modulated light source}

\author{Z.~L.~Yuan}
\email{zhiliang.yuan@crl.toshiba.co.uk}
\author{B.~Fr\"ohlich}
\author{M.~Lucamarini}
\affiliation{Toshiba Research Europe Ltd, 208 Cambridge Science Park, Milton Road, Cambridge, CB4~0GZ, UK}

\author {G.~L.~Roberts}
\affiliation{Toshiba Research Europe Ltd, 208 Cambridge Science Park, Milton Road, Cambridge, CB4~0GZ, UK}
\affiliation{Cambridge University Engineering Department, 9 J J Thomson Ave., Cambridge, CB3 0FA, UK}

\author {J.~F.~Dynes}
\author {A.~J.~Shields}
\affiliation{Toshiba Research Europe Ltd, 208 Cambridge Science Park, Milton Road, Cambridge, CB4~0GZ, UK}

\date{\today}

\begin{abstract}

The art of imparting information onto a light wave by optical signal modulation is fundamental
to all forms of optical communication. Among many schemes, direct modulation of laser
diodes stands out as a simple, robust, and cost effective method. However, the simultaneous
changes in intensity, frequency and phase are a drawback which has prevented its application in the field of secure quantum communication.
Here, we propose and experimentally demonstrate a directly phase-modulated light source which overcomes the
main disadvantages associated with direct modulation and is suitable for diverse applications
such as coherent communications and quantum cryptography. The source separates
the tasks of phase preparation and pulse generation between a pair of semiconductor lasers
leading to very pure phase states. Moreover, the cavity enhanced electro-optic effect enables the first example of
sub-volt halfwave phase modulation at high signal rates. The source is compact, stable and
versatile, and we show its potential to become the standard transmitter for future quantum communication networks.
\end{abstract}

\maketitle

Phase modulation is an important encoding format\cite{agrawal02} and forms the basic building block for other signal formats such as amplitude, polarisation \cite{tittel2009}, and quadrature amplitude modulation \cite{winzer2006}.
The primary enabling technology is external modulation, which uses electro-optic materials whose refractive index varies with electric-field \cite{saleh1991fundamentals}. Conventional phase modulators based on LiNbO$_3$ crystals \cite{noguchi98} or semiconductor waveguides \cite{smit2014} require driving voltages beyond the reach of complementary metal-oxide semiconductor (CMOS) logic, necessitating the use of amplifiers which add to the system cost and complexity. Their prospects for a sub-volt halfwave voltage ($V_\pi$) suitable for direct CMOS driving are limited. Substantially increasing the device length is both undesirable and difficult. LiNbO$_3$ phase modulators already possess a length of several centimeters, while semiconductor devices suffer from impedance matching problems \cite{ding2011jlt} and considerable insertion loss at longer lengths.
Organic dielectric materials show promising electro-optic coefficients \cite{shi2000science}, but as yet have not resulted in a sub-volt phase modulator \cite{enami2007}.

In quantum key distribution (QKD), which is a powerful method for protecting future communication networks \cite{gisin02,lo2014secure}, the most common way to transmit quantum information is as an optical phase delay between weak coherent pulses, as this has been shown to be robust even for transmission over installed fiber networks \cite{peev09,sasaki11,wangOE14,tang2016prx}.  For the well-known BB84 protocol, the phase encoded pulses are usually generated by passing a laser pulse through an asymmetric Mach-Zehnder interferometer, requiring several discrete free space or fiber optical devices and components \cite{townsend1994secure,hughes2000quantum}. This has resulted in a structurally cumbersome transmitter that needs active stabilisation \cite{yuan2005continuous} in addition to requiring powerful driving electronics for phase modulation.  We note that such BB84 encoder is incompatible with other QKD protocols, such as differential-phase-shift \cite{inoue2002} and coherent-one-way \cite{stucki2005fast}.

Direct modulation of laser diodes is attractive because no external modulator is necessary, and it can be achieved with low drive voltages \cite{yamamoto2012}.
However, spurious effects such as frequency chirp, large intensity fluctuations, and uncontrolled phase evolution limit its applicability.
To date, it has been used primarily for on-off keying in classical communications with restricted data bandwidth and/or communication distances \cite{winzer2006}.
Although it also produces phase modulation, the dominant amplitude variation makes it non-ideal for state-of-the-art coherent communication systems \cite{cano2014direct}. In challenging applications like quantum cryptography, direct modulation fails altogether, as the unintentional changes in other degrees of freedom leak side-channel information \cite{scarani2014black}.

Here, we introduce a novel concept for a directly phase-modulated light source.
Counterintuitively, it employs only laser diodes as active components, but operates on the principle of external modulation by using one diode solely as an electro-optic device for phase control.
The resulting source overcomes disadvantages of direct modulation, while retaining all benefits associated with this technique.
It features an exceptionally low drive voltage, excellent phase stability, and great versatility, making it an attractive choice for many applications, including quantum cryptography, which we will demonstrate.

Figure~1(a) shows a schematic of our phase-modulated light source, consisting of a pair of laser diodes connected via an optical circulator.
The phase preparation laser is directly modulated to produce a train of nanosecond duration, quasi steady-state emission. Each of these long pulses seeds coherently a block of two (or more) secondary, short optical pulses ($<$100 ps) from the pulse
generation laser. The relative phase of these secondary pulses can be set to an arbitrary value by directly modulating the driving current applied to the phase preparation laser, while their intensity and frequency are essentially unaffected.

An intuitive picture helps to understand how we prepare an optical phase, Figs.~1(b) and (c).
Consider a steady-state laser with its optical phase evolving at a constant rate of $2\pi\nu_0$, where $\nu_0$ is its central optical frequency.
Under a small perturbation, the optical frequency shifts by an amount $\Delta \nu$, changing the course of the phase evolution.  This perturbation will create a phase difference,
\begin{equation}\label{e}
 \Delta \phi = 2\pi\Delta\nu t_m,
\end{equation}
\noindent where $t_m$ is the duration of the perturbation.
The perturbation signal here is an electrical modulation to the phase preparation laser, and the optical frequency change arises from the
effect of the carrier density on the refractive index in the active laser medium\cite{bennett90}.
Unlike in existing 
modulators, the phase shift in Eq. (1) depends on the duration of the modulation signal.
We attribute this dependence to cavity enhancement: the laser cavity confinement allows the light field to oscillate back and forth within the cavity and thus experience the refractive-index change of the active medium over the entire duration of a modulation signal.  As shown later, this cavity effect has allowed to realise the first sub-volt $V_\pi$ phase modulation.

We use an asymmetric Mach-Zehnder interferometer  (Fig.~2(a)) to measure the relative phases of a train of short pulses at 500~ps intervals from the pulse generation laser. The interval is sufficiently long that the field within the pulse generation laser is extinguished between pulses, such that lasing can be triggered by either spontaneous emission \cite{yuan14}  (with random phase) or the phase preparation laser (with defined phase). Figure~2(a) (right-hand panel) compares the case where the pulse generation laser is seeded with light from the phase preparation laser to the case where it is unseeded. The unseeded case produces output waveforms of random intensities, while an injection of continuous-wave light leads to a fixed phase difference resulting in a stable output intensity.  The fidelity of the phase transfer between the laser diodes is evaluated by the interference visibility of the short pulses, which is found to grow with the injection strength and saturates at 99.06\% with a modest injection power of 50~$\mu$W.

We now demonstrate phase modulation by applying a small perturbation pattern to the electrical drive of the phase preparation laser, which produces a shallow intensity variation, Fig.~2(b)(i). The key point to note here is that the perturbation does not disrupt the phase continuity, but only changes the phase evolution rate. We make use of this change by synchronising each modulated signal to the interval of a pair of the output pulses whose relative phase is to be modified.
This arrangement ensures also the indistinguishability among the output pulses, because they will all be seeded by the unmodulated part of the injected light. The successful transfer of the electrical drive pattern to the output phase is confirmed by the resulting interference waveforms, which reproduce the modulation pattern, Fig.~2(b)(ii) and (iii).

Figure~2(c) shows the phase shift measured as a function of the modulation voltage to the phase preparation laser.  
The phase shift can be either positive or negative, and is approximately linear with the signal amplitude.
We determine a halfwave voltage of 0.35~V. 
We ascribe this low $V_\pi$ to the aforementioned cavity enhancement, which enables an effective interaction distance of 25~mm despite the active medium having a length of only $\sim$100$\mu$m.
As the $V_\pi$ is sufficiently low to be driven directly by CMOS logic, we expect this breakthrough will significantly reduce the complexity, as well as energy and physical footprint, of a phase-modulated light source.

To highlight the versatility of the source and to demonstrate its practicality as a QKD light source, we introduce phase randomisation by modifying the driving signal to the phase preparation laser.
The driving signal is set below the lasing threshold for 250~ps during each 1~ns period to stop the light emission, Fig.~3(a).  The depletion of the light field forces the next lasing period to be triggered by the vacuum fluctuation and hence with random phase relative to the previous pulse.
Adjacent short pulses seeded by different quasi steady-state pulses will therefore have a random phase as shown in Fig.~3(b), whereas pulse pairs by same seed pulse show phase coherence. Our source therefore meets the requirement for global phase randomisation required for the security of the BB84 protocol \cite{BB84}.

We integrate the source in a BB84 transmitter to demonstrate its suitability for QKD applications.
Figure~4(a) shows the results, where the sifted key rate and quantum bit error rate (QBER) are directly measured quantities.
The experimental values (symbols) are in excellent agreement with theoretical simulation (lines).  The maximum transmission loss of $\sim$40~dB (equivalent to 200~km of standard fiber) is limited by the detector noise.  The QBER stays approximately constant at a base level of 2.4\% for channel losses up to 30~dB. This base value sets an upper bound for the encoding error of the light source as a BB84 encoder, which is comparable to the values achieved with conventional bulk or fiber optics \cite{lucamarini13OpEx}.

Owing to its interferometer-free design, the phase-modulated source has excellent phase stability. This reduces the complexity of the QKD setup by removing the need for active stabilization of the phase.
For illustration, we measure the QBER continuously over a 24-h period with no active feedback applied.
The light source, and the receiving interferometer, are independently temperature-controlled. The time-dependent QBER, sampled at an interval of 1~s, fluctuates within a very narrow range around an average value of 2.41\% with a standard deviation of 0.10\%, Fig.~4(c).  We plot the distribution of the measured QBER in the inset, together with a simulation of the unavoidable shot-noise statistical fluctuation due to the finite count rate.
The close resemblance between the two distributions suggests that the additional phase instability due to the light source is insignificant.

By simply applying different electrical signals, the light source can be reconfigured to accommodate a variety of QKD protocols, including differential-phase-shift \cite{inoue2002} and its recent ``round-robin" variant \cite{sasaki2014nature}. It is also possible to implement the coherent-one-way protocol \cite{stucki2005fast} by introducing a binary pattern in the driving signal to the pulse generation laser.
To demonstrate this reconfigurability, we use the same BB84 optical setup to implement the differential-phase-shift protocol.
Figure~4(b) shows the experimental results (symbols) together with the theoretical simulation (lines). The base QBER of 1.9\% is well within the error threshold of the protocol.
We measure also the performance over a 100~km fiber spool, observing very similar error and bit rates to using the equivalent optical attenuation shown by the grey data points in the plot.

We have demonstrated a novel directly modulated light source which permits to prepare very pure phase states with exceptionally low driving voltage and which is suitable for challenging applications such as QKD.
The phase-modulated source could be integrated into a fully-functional phase transmitter with a size comparable to small-form pluggable transceiver modules (SFPs) ubiquitously found in today's communication systems. Integration at this level is highly desirable, and will find applications in heterogeneous networks where different quantum communication protocols coexist \cite{sasaki11} or in access networks where the compactness and cost of the transmitters is of paramount importance \cite{frohlich2013nature}.
Beyond communication applications, the direct source is useful to provide a stable phase-conditioned pulse sequence for the control of quantum systems, such as quantum dots \cite{jayakumar2014} and parametric down-conversion \cite{de2002}. 

\section*{Appendix}

\subsection{Directly phase-modulated light source.}
Fiber-pigtailed distributed feedback laser diodes with built-in coolers are used for the setup shown in Fig.~\textbf{1}. The pulse generation laser is biased just above its lasing threshold.  In the absence of optical injection it has an output power of $\sim$20~$\mu$W.
We temperature-tune the wavelength of the phase preparation laser for resonant injection, resulting in a resonant enhancement in the intensity of the pulse generation laser. The optical power output of the pulse generation laser increases to 275~$\mu$W in the presence of 50~$\mu$W injection from the phase preparation laser.
The direct source emits 70~ps pulses at 1551~nm with a repetition rate of 2~GHz.

\subsection{Phase measurements.}
A planar lightwave circuit based Mach-Zehnder interferometer (Fig.~2(a)) is used to
evaluate the relative phase between adjacent pulses from the light source. The power splitting ratios of both input and output beam-splitters are nominally 50/50. The differential delay is 500~ps, and a built-in heater is used to adjust the interferometer phase. An optical powermeter or an oscilloscope records the interference results, which depend on whether a modulation is applied to the phase preparation laser. Under continuous-wave injection, the measured visibility of 99.02\% of the output pulses is limited by the coherence of the phase preparation laser diode we used in this experiment. A visibility of 99.92\% has been observed when we replace the laser diode with a laser with longer coherence time (spectral width: 150~kHz).
To measure the data in Fig.~2(c) we apply a fixed modulation pattern to the phase preparation laser to enable the direct source to produce a train of short pulses with a phase pattern of ``0 0 $\Delta\phi$ $\Delta\phi$".   By varying the interferometer phase, we obtain two distinctive interference fringes corresponding to ``0" and ``$\Delta \phi$''  phase, respectively. For each signal amplitude, we determine the $\Delta \phi$ value by comparing the interference fringes.

\subsection{Quantum key distribution experiments.}
The phase-modulated source transmits phase encoded light pulses at a clock rate of 2~GHz, leading to effective QKD clock rates of 1 and 2~GHz for the BB84 and differential-phase-shift protocols, respectively.  We use a quantum random number generator \cite{yuan14} to produce a 256-symbol sequence without intentional bias for each protocol. Modulation is applied to each pulse pair with a differential phase delay among [0, $\pi/2$, $\pi$ and $3\pi/2$] for the BB84 protocol, while each pulse is set to either 0 or $\pi$ phase in relation with its preceding one in the differential-phase-shift protocol.
The intensity of the source is heavily attenuated to the respective single photon levels, 0.5 and 0.4 photons/ns.  An optical attenuator simulates the fiber channel with a loss scaling rate of 0.2~dB/km.  Temperature-controlled planar lightwave circuit Mach-Zehnder interferometers of 3~dB loss are used for phase decoding, and the decoding basis is chosen passively with a 50/50 beam-splitter for the BB84 protocol.
For single photon detection, we use a superconducting nanowire detector system featuring a detection efficiency of 13--15\% and a dark count rate of 150~Hz at a wavelength of 1550~nm. Time-tagging electronics record photon detection events, from which we apply a 0.25~ns detection window every 0.5~ns to extract sifted key rates and QBER's.
To compute the secure key rates, we follow the equations in Refs \cite{lucamarini13OpEx} and \cite{shibata2014} for the decoyed BB84 and differential-phase-shift protocols, respectively.

\acknowledgments
We thank J. Huwer for helping setting up the superconducting detectors. G. L. R. acknowledges personal support via the EPSRC funded CDT in Photonic Systems Development and Toshiba Research Europe Ltd.



%

\newpage
\begin{figure}
\centering
\includegraphics[width=.85\columnwidth]{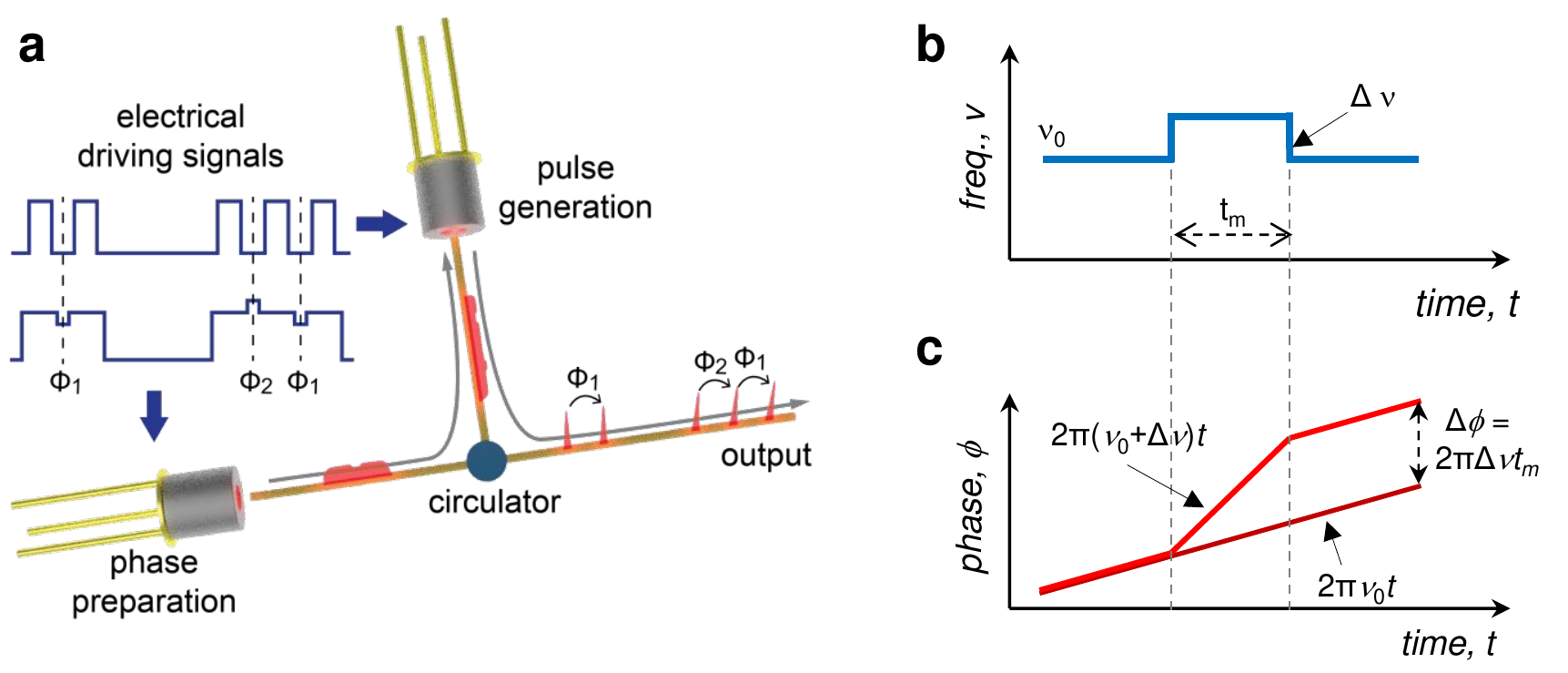}
\caption{Directly phase-modulated light source. (a) The source consists of a pair of semiconductor laser diodes connected via an optical circulator.  We refer to these laser diodes as the phase preparation and pulse generation lasers.
The phase preparation laser is biased to produce nanosecond scale, quasi steady-state optical pulses with shallow intensity modulation which modifies also the optical phase.  The gain-switched pulse generation laser emits short optical pulses which inherit the optical phase prepared by the phase preparation laser.
The duration of each seed pulse can be varied to seed a pulse train of different lengths.
(b) The optical frequency of a steady-state laser under a small perturbation of duration $t_m$.  (c) Optical phase trajectories with and without the perturbation.}
\label{fig:fig1}
\end{figure}

\begin{figure}
\newpage
\centering
\includegraphics[width=.96\columnwidth]{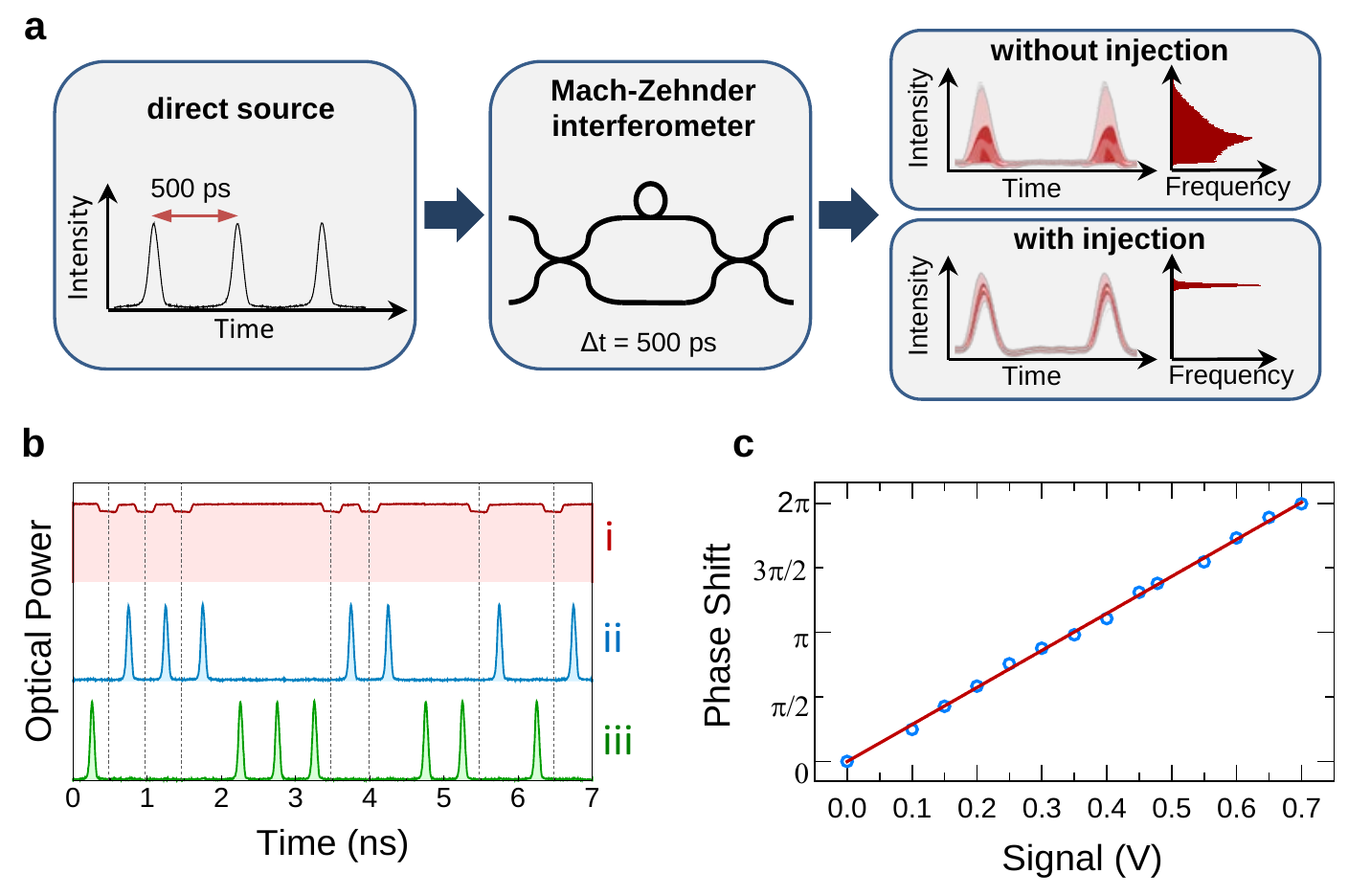}
\caption{Coherence seeding and phase modulation.
(a) Experimental setup for evaluation of the relative phase between adjacent short pulses from the directly phase-modulated light source:  The source transmits a pulse train of 500~ps intervals and an asymmetric Mach-Zehnder interferometer (MZI) with a matching delay is used to measure the interference. Panels on the right show colour coded density plots of the measured waveforms and corresponding histograms of peak intensities for the cases with and without optical injection, respectively.
(b) Demonstration of direct phase modulation; (i) Output intensity profile of the phase preparation laser with shallow intensity modulation.
(ii), (iii) Complementary intensity profiles recorded from the output of the MZI.  (c) Phase shift as a function of the electrical signal amplitude applied to the phase preparation laser.  In (b) and (c), the modulation signal has a duration of 250~ps.
}
\label{fig:fig2}
\end{figure}

\begin{figure}
\newpage
\centering
\includegraphics[width=.6\columnwidth]{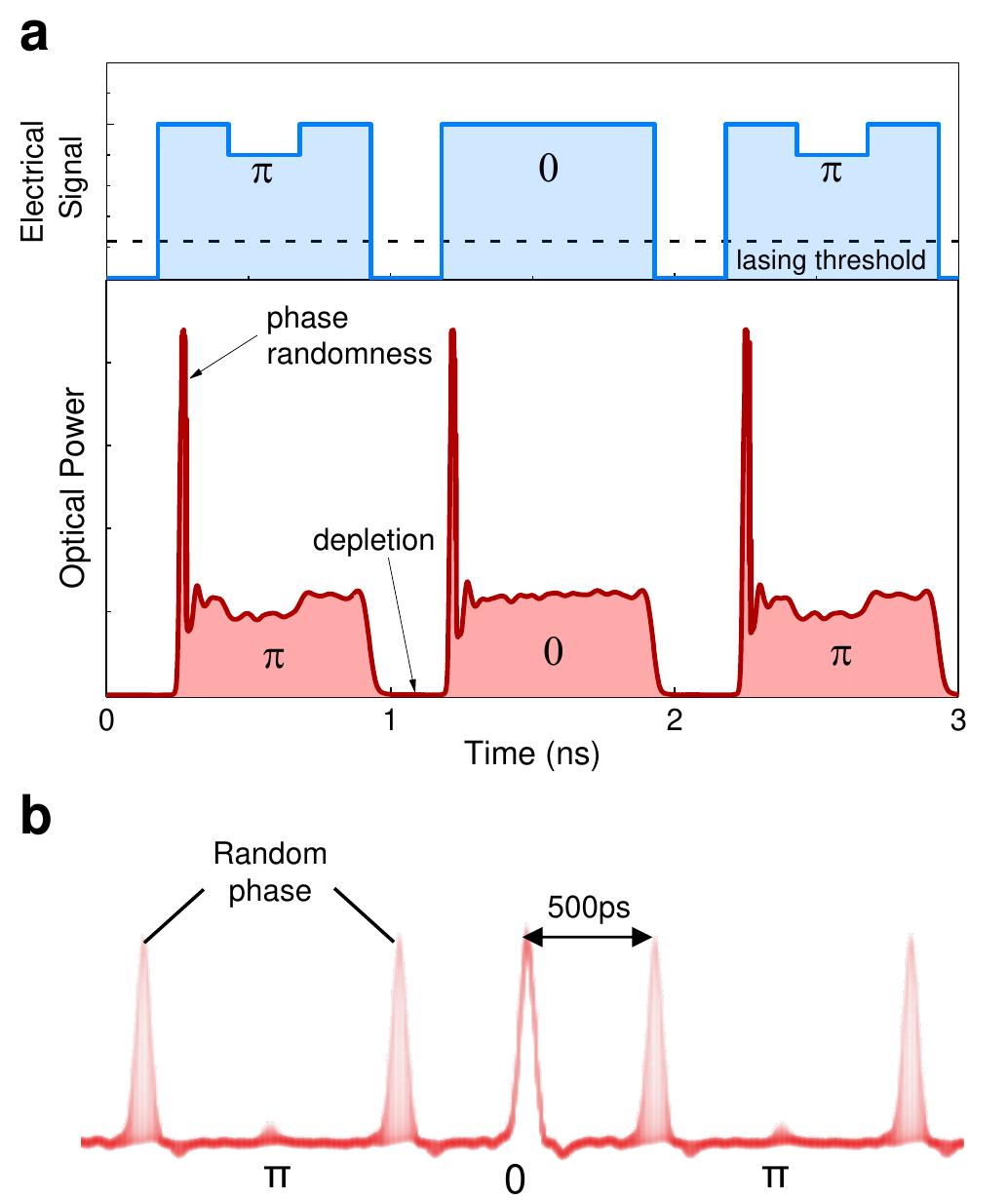}
\caption{Phase randomisation is required for the security of the BB84\cite{BB84} and ``round-robin" differential phase shift\cite{sasaki2014nature} protocol. We realise it here by depleting the cavity photon field in the phase preparation laser prior between emission periods.  (a) Upper panel: The driving signal applied to the phase preparation laser;  Lower panel: Temporal emission profile of the phase preparation laser;
The profile shows relaxation oscillations at the start of each pulse which quickly settles into steady state emission. Relaxation oscillations are caused by an initial overshoot when emission start from a depleted cavity. The duration of each quasi steady-state emission is long enough to coherently seed a pair of short output pulses.
(b) Colour coded density plot of the measured waveform of short pulses seeded by the injected light with a pattern shown in (b).  The outputs with well-defined (random) intensity correspond to the interference of short pulses seeded by same (different) seed pulse.
}
\label{fig:fig1}
\end{figure}

\begin{figure}
\newpage
\centering
\includegraphics[width=0.8\columnwidth]{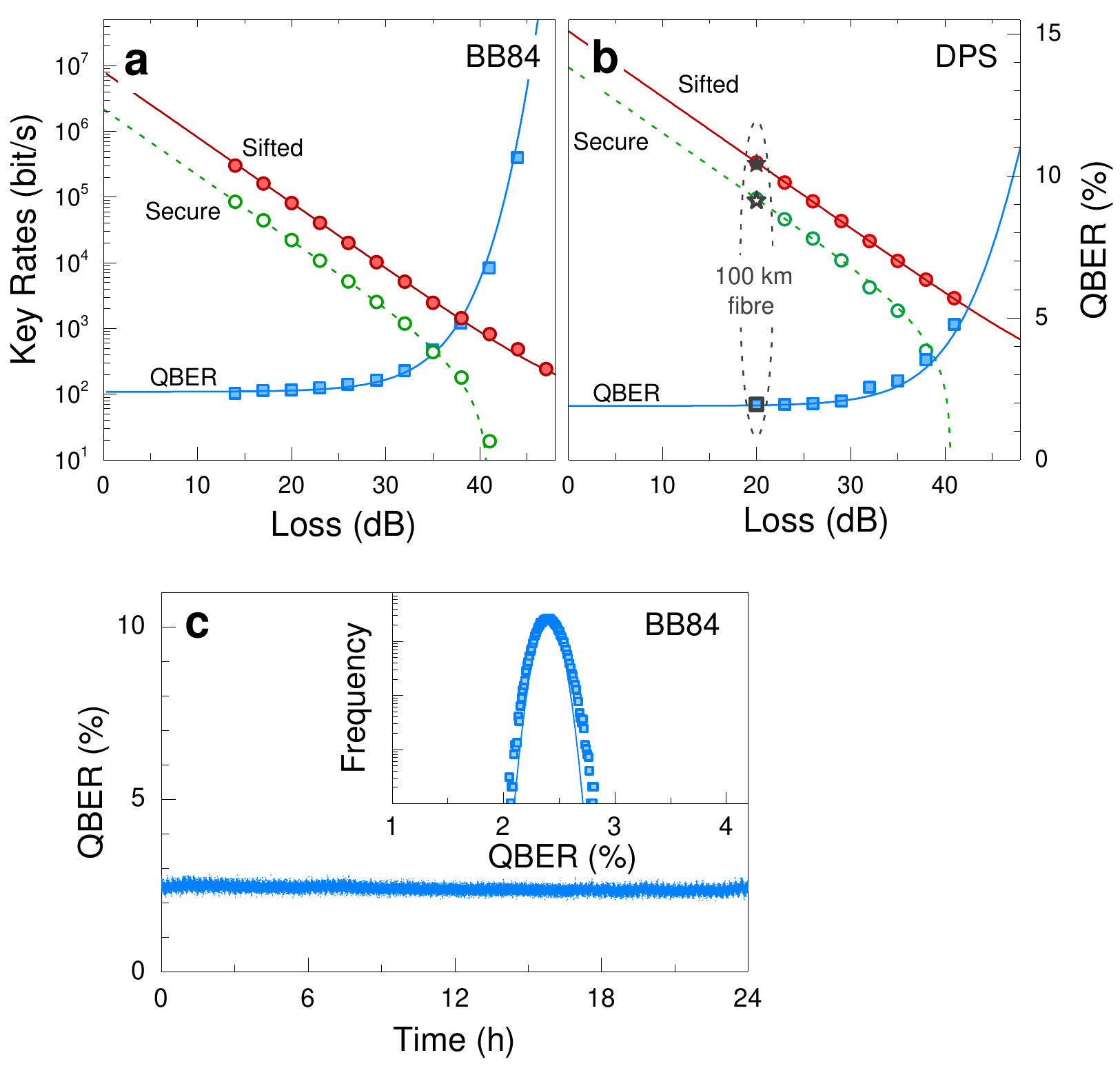}
\caption{QKD operating with the directly phase-modulated light source. (a) BB84 protocol: sifted and secure rates, and QBER versus channel loss; (b) QKD results using the differential-phase-shift protocol.
(c) QBER measured with the BB84 protocol for a duration of 24 hours. The inset shows the distribution of the measured QBER's (symbols) together with a simulation (line) of the shot-noise statistical fluctuation from the finite count rate; The signal integration time is 1~s.
}
\label{fig:qkd}
\end{figure}

\end{document}